# An Initial Step Towards Organ Transplantation Based on GitHub Repository

**Shangwen Wang[1], Xiaoguang Mao[1], and Yue Yu[1]**
[1]College of Computer Science, National University of Defense Technology, Changsha, 471003, China

Corresponding author: Xiaoguang Mao (e-mail: xgmao@nudt.edu.cn).

This work was funded by the National Natural Science Foundation of China (Grant No.61379054, 61672592).

**ABSTRACT** Organ transplantation, which is the utilization of codes directly related to some specific functionalities to complete one's own program, provides more convenience for developers than traditional component reuse. However, recent techniques are challenged with the lack of organs for transplantation. Hence, we conduct an empirical study on extracting organs from GitHub repository to explore transplantation based on large-scale dataset. We analyze statistics from 12 representative GitHub projects and get the conclusion that 1) there are abundant practical organs existing in commits with "add" as a key word in the comments; 2) organs in this repository mainly possess four kinds of contents; 3) approximately 70% of the organs are easy-to-transplant. Implementing our transplantation strategy for different kinds of organs, we manually extract 30 organs in three different programming languages, namely Java, Python, and C, and make unit tests for them utilizing four testing tools (two for Java, one for Python, and one for C). At last, we transplant three Java organs into a specific platform for a performance check to verify whether they can work well in the new system. All the 30 organs extracted by our strategy possess good performances in unit test with the highest passing rate reaching 97% and the lowest one still passing 80% and the three Java organs work well in the new system, providing three new functionalities for the host. All the results indicate the feasibility of organ transplantation based on open-source repository, bringing new idea for code reuse.

## I. INTRODUCTION

With the increasing number of software being developed, several engineers proposed about extending the functionality of their individual software by facilitating others' codes. This method is called code reuse, which has been extensively studied [11]. For example, software component, referred to a set of classes, is considered as the basic unit for reuse [23].

In 2015, Harman et al. [13] proposed a new concept, organ, which refers to all codes associated with the feature of interest, bringing a new chance for software reuse. Different from components underlining high relationship between several classes, organs emphasize on the integrity on functionality. Organs do not have to be several classes. It can be several lines of code, a function, or one class, as long as it finishes a specific functionality independently. If one function alone can fulfill the needs of a software whereas a set of classes is added into the software, it is then highly time-consuming to remove the extra codes, not to mention the defects which might be induced by the redundancy of codes. This redundancy problem would be resolved by retrieval of the functional codes and the corresponding transplantation of them into the target software. Hence, the overburdened code transplantation and its negative impact is expectedly avoided. That is why organ is a more flexible unit for reuse and brings much more convenience than traditional software reuse based on component. However, the practice [14, 15] in this area is restricted to a small-scale and specific experimental context. The general exploration relating to organ extraction and transplantation based on the large-scale dataset has not been well studied.

Open-source movement is becoming popular recently [26]. Many developers are joining the collaborative development community to develop projects iteratively. These developers continue committing their codes to the repository. Thus, the repository keeps track of the progress of a project. Those codes which are added into the repository by contributors may contain some functionalities that are remarkable to other developers. Whether we can obtain several practical organs for transplantation through complete analysis of repository remains unknown. New ideas and direction will be brought to researchers if it is feasible.

On this basis, we put forward a strategy for transplanting organs from repositories and present an empirical study on extracting practical organs from GitHub repository, aiming at remedying the problem of lacking of organs. The main contributions of this paper are:

- We divide commits in GitHub into eight categories based on the keywords in the comments and get the conclusion that abundant practical organs are in the adding commits.
- We find that there are four kinds of common contents in the organs and calculate their percentages.
- We define our criteria for dividing organs into two categories: easy-to-transplant and difficult-to-transplant and then display the statistics which show the percentages for both kind.
- We put forward a strategy for extracting and transplanting organs from open-source repositories for both types of organs (easy-to-transplant and difficult-to-transplant).
- We conduct an empirical study on extracting and transplanting organs using our methodology and the feasibility of our approach has been proved.

The remaining part of this paper is organized as follows. Section Ⅱ introduces the design of our study. Section Ⅲ provides the answers to each research question in our study and makes detailed analyses. Section Ⅳ begins by presenting the discussion about the threats to validity in our study. Section Ⅴ provides a review on projects and studies related to this topic. Section Ⅵ presents the conclusion to our study and our future work.

## II. STUDY DESIGN

The study design discusses our research questions and datasets, aiming at ensuring that the design is appropriate for the objectives of the study.

### A. RESEARCH QUESTIONS

In particular, we address the following research questions in our study:

**RQ1:** *Is there any evident access of organ extraction?*

The bulk of information in the open-source repository adds to the difficulty in locating the practical organs accurately and efficiently. To answer this question, we analyze 12 projects from GitHub repository to find whether there existing evident symbols for extraction. The answer to this question will provide reference for locating practical organ, thereby making preparation for automatic extraction of organs in the future.

**RQ2:** *How many kinds of contents do these organs contain?*

To answer this question, we divide organs into different types based on their contents. The answer will reveal the portion of the organs with different contents and make response to questions about organ abundance.

**RQ3:** *What is the portion of easy-to-transplant organs?*

To answer this question, we define the concepts of easy-to-transplant and difficult-to-transplant organs based on the degree of relevance of the organ with the source codes, which will provide a theoretical basis for extraction method.

**RQ4:** *How to transplant these organs?*

To answer this question, we put forward a strategy for transplantation which takes different operations when facing different kinds of organs: for easy-to-transplant organs which means they have weak relevance with the source code, we transplant them directly; for difficult-to-transplant organs possessing strong relevance with the source code, we take some special operations to make preparation for their being transplanted.

**RQ5:** *Do the organs extracted from GitHub repository work well?*

To answer this question, we extract organs under the instruction of our strategy and check the performances for them. We first extract organs in different programming languages and make unit tests for them. Then we transplant 3 organs into a software on the market to see whether this software possesses special functionalities after transplantation. The answer will verify whether our topic can be implemented in the automated transplantation field in the future.

### B. DATASETS

For RQ1 to RQ3, our target is to make an empirical study and get the locations of potential organs, the contents of organs, and the portion of easy-to-transplant organs. Given the limitation of time and manual effort, we determine to choose twelve popular projects to conduct this empirical study, meaning that each project must be highly representative. Our intuition is that a project is likely to be more popular and representative if it is collected by more users and thus we choose number of collectors as the criterion. The GitHub Explore home page [4] presents six kinds of collection, each of which shows a list of projects related to this topic. We first select 10 projects from these six topics, each of them possessing more than 300 collectors. Therefore, these projects, including *decisiontree* [1] and *awesome-nlp* [2] from *Getting started with machine learning*, *SoundJS* [3] from *Music*, *Stethoscope*[4] and *polygons*[5] from *Social Impact*, *lightcrawler*[6] and *a11y*[7] from *Web accessibility*, *lint*[8] from *Clean code linters*, and *square.github.io*[9] and *twitter.github.com*[10] from *GitHub Pages examples*, are relatively popular in related fields. However, none of these projects is written in Java, the most popular programming language in GitHub. We then decide to add two projects written in Java, *requery* [11] and *FabulousFilter*[12] which are collected by 2.5k and 1.8k users respectively, into our dataset to make our conclusion more general. The distribution of the programming languages in our dataset is shown in Table I. Note that *awesome-nlp* is a library for machine learning algorithms and it just provides references, thus it does not use any language. Statistics show that projects in our dataset are written by six of the most popular programing languages in the world. We collect this dataset on July 11, 2018. All the changes in repository that follow this date are excluded from our study.

[1]https://github.com/igrigorik/decisiontree
[2]https://github.com/keon/awesome-nlp
[3]https://github.com/CreateJS/SoundJS
[4]https://github.com/GliaX/Stethoscope
[5]https://github.com/ncase/polygons
[6]https://github.com/github/lightcrawler
[7]https://github.com/addyosmani/a11y
[8]https://github.com/golang/lint
[9]https://github.com/square/square.github.io
[10]https://github.com/twitter/twitter.github.com

TABLE I
DISTRIBUTION OF PROGRAMING LANGUAGES IN OUR DATASET

| Programing Language | Project |
|---|---|
| java | requery & FabulousFilter |
| go | lint |
| javascript | lightcrawler & twitter.github.io & square.github.io & SoundJS |
| css | polygons |
| ruby | Stethoscope & dicisiontree |
| html | a11y |

For RQ5, we extract ten organs written in Java from [5], ten organs written in Python from [7], and ten organs written in C from [9] to make unit tests. Then we extract three Java organs with real-world functions from three Java corpuses [1, 2, 3] and transplant them into a Java real-world application. To the best of our knowledge, this dataset, containing thirty-three real-world organs, is the largest one from open-source repository for transplantation study so far and it is different from datasets used by state-of-the-art tools like mu_scalpel [14]: it is extracted from codes added by developers in open-source environment while previous ones are from ready-made systems or software.

## III. STUDY RESULTS

In this section, we analyze information from GitHub repository, propose our policy for transplantation, and conduct experiments to answer the research questions.

### A. RQ1: IS THERE ANY EVIDENT ACCESS OF ORGAN EXTRACTION?

A project is incompletely functional at its inception, and many of its features are later added stepwise by engineers, meaning that we can extract organs with remarkable functionality from the codes added by developers in a project's repository. We focus on the commits based on the above analysis to study whether an evident indication, which suggests that commits are about addition of some codes, exists. The intuition is that if the commits are about adding some codes, it may possess the functionality we would like to utilize and thus an organ exists in these codes.

When committing the codes, a contributor often writes comments on the code operation done, to reveal his/her intention to change the codes. When a bug is repaired, comments would be "fix a certain error" or "correct a certain bug". "Add" is used for some new functionalities added to the project and "remove" or "delete" is used when some original files are deleted. Replacing the old resource by a neoteric one is then commented as "update", creating a new file in this project as "create". The word "modify" or "change" is to label the work done for modifying the project. Likewise, a comment often begins with "merge" when a developer decides to merge a pull request when identifying something practical from it from a branch of this project. According to this finding, we divided the commits into eight categories based on the keywords in the comments. Each category displays one or two keywords, which are update, fix/correct, add, delete/remove, modify/change, merge, create, and other, respectively.

We calculate the number of occurrences of each category of keywords in each project and display the results in Table II. Commits using add as a keyword become our study objects according to our intuition, since they indicate the existence of potential organs. Nevertheless, organs may still exist in commits using other keywords like "fix". Due to the limitation of time and human-effort, a deeper study is in our future work. Commits using add as a keyword in comment occupy more than 20% of the total amount of commits, which indicates that we can obtain information about the functionality added to the project by locating these commits. We call these commits *adding commits*. Note that in some rows, for example, the SoundJS, the sum of each number of the occurrence of each category is larger than the total amount of commits. That is caused by the *multi-keyword* phenomenon illustrated in Figure 1: there may be more than one keyword in a comment. On this occasion, the number of occurrence of each corresponding category is added by one.

Nevertheless, uncertainty still exists about the practicality of these commits. Whether these commits provide practical organs remains unknown. We define a standard to evaluate the practicality of an organ through these four conditions: (1) adding some explanations, icons, and pictures into the projects, (2) adding a missing symbol into the codes, (3) adding dependencies into the configuration files, and (4) adding some words into README file as annotation (examples are shown in Figure 2 to Figure 5). We consider these four conditions unpractical because we cannot extract organs with functionality that we are interested in. We call these commits as unpractical adding. Except from these two conditions, other commits using add as a keyword add new functionalities into the projects according to our empirical observation. These codes can be facilitated by other engineers. We consider these commits as practical adding. We calculate the ratio of practical adding to the total amount of adding commits in each project, and the result is shown in Table III.

Nearly 60% of *adding commits* are practical according to the result which means the majority of adding commits have practical functionalities, indicating that we can determine practical organs from most adding commits.

[11] https://github.com/requery/requery

[12] https://github.com/Krupen/FabulousFilter

**Answer to RQ1:**

We can consider commits using add as a keyword in the comments as evident symbols for extraction. Most of these commits provide practical organs, which possess functionalities that other engineers might be interested in.

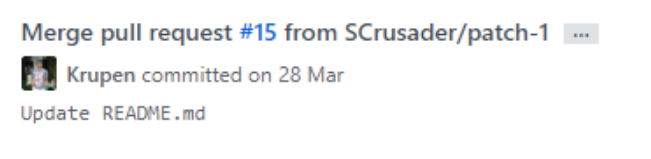

**FIGURE 1. A case of multi-keyword phenomenon. Note that the keywords "merge" and "update" are both in this comment.**

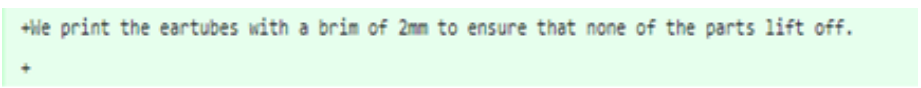

**FIGURE 2. A case of adding explanation into the project**

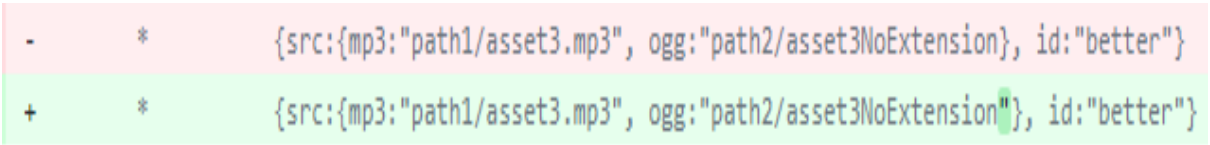

**FIGURE 3. A case of adding a missing symbol into the project**

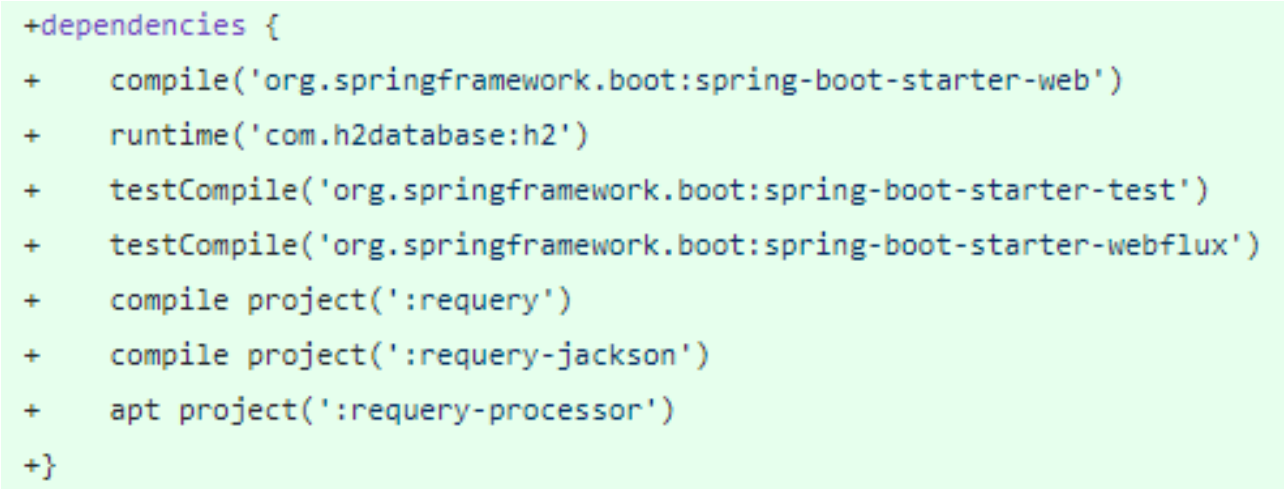

**FIGURE 4. A case of adding dependency into configuration file**

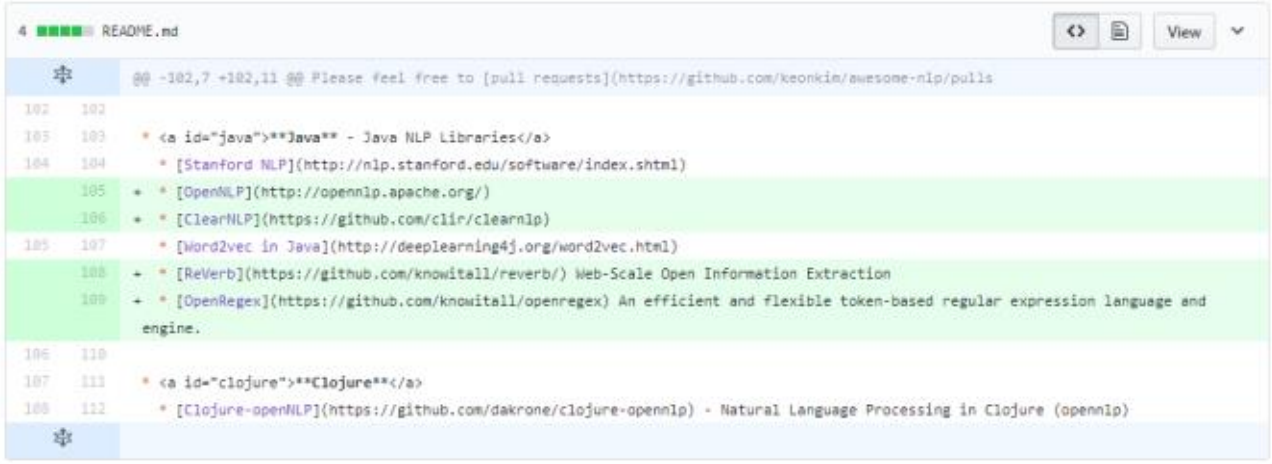

**FIGURE 5. A case of adding annotation into readme file**

TABLE II
NUMBER OF OCCURRENCES OF EACH CATEGORY IN THE 12 PROJECTS AND THEIR PERCENTAGES

| **Project Name** | *update* | *fix/correct* | *add* | *delete/remove* | *modify/change* | *merge* | *create* | *other* | **Total amount of commits** |
|---|---|---|---|---|---|---|---|---|---|
| SoundJS | 240 | 168 | 98 | 58 | 41 | 25 | 24 | 117 | 602 |
| decisiontree | 11 | 14 | 11 | 4 | 3 | 18 | 2 | 20 | 69 |
| awesome-nlp | 77 | 24 | 112 | 21 | 8 | 99 | 13 | 10 | 269 |
| Stethoscope | 18 | 16 | 44 | 14 | 8 | 5 | 4 | 59 | 159 |
| polygons | 0 | 10 | 13 | 0 | 4 | 1 | 0 | 92 | 111 |
| lightcrawler | 1 | 0 | 5 | 1 | 0 | 1 | 0 | 19 | 27 |
| a11y | 15 | 15 | 53 | 5 | 2 | 17 | 1 | 83 | 162 |
| lint | 22 | 61 | 49 | 8 | 18 | 0 | 0 | 83 | 167 |
| square.github.io | 9 | 8 | 34 | 10 | 0 | 25 | 0 | 23 | 79 |
| twitter.github.com | 12 | 13 | 16 | 2 | 3 | 16 | 0 | 21 | 65 |
| requery | 199 | 181 | 131 | 63 | 19 | 4 | 19 | 217 | 816 |
| FabulousFilter | 22 | 11 | 5 | 6 | 0 | 3 | 1 | 6 | 54 |
| **In total** | 626 | 521 | 571 | 192 | 106 | 214 | 64 | 750 | 2,580 |
| **Percentage(%)** | 24.26 | 20.19 | 22.13 | 7.44 | 4.11 | 8.29 | 2.48 | 29.07 | |

TABLE III
NUMBER OF OCCURRENCES OF PRACTICAL AND UNPRACTICAL ADDING IN EACH PROJECT AND THEIR PERCENTAGES

| **Project Name** | *practical adding* | *unpractical adding* | **Total amount of *adding commits*** |
|---|---|---|---|
| SoundJS | 62 | 36 | 98 |
| decisiontree | 9 | 2 | 11 |
| awesome-nlp | 33 | 79 | 112 |
| Stethoscope | 19 | 25 | 44 |
| polygons | 7 | 6 | 13 |
| lightcrawler | 2 | 3 | 5 |
| a11y | 35 | 18 | 53 |
| lint | 40 | 9 | 49 |
| square.github.io | 16 | 18 | 34 |
| twitter.github.com | 11 | 5 | 16 |
| requery | 102 | 25 | 131 |
| FabulousFilter | 1 | 4 | 5 |
| **In total** | 337 | 234 | 571 |
| **Percentage(%)** | 59.02 | 40.98 | 100.00 |

### *B. RQ2: HOW MANY KINDS OF CONTENTS DO THESE ORGANS CONTAIN?*

We examine the codes added by engineers to answer this question carefully. Several kinds of codes may be added by contributors when they intend to satisfy some functionalities. Contributors may add one or more statements or a new function to the program. A new class is also defined when the new functionality is, to some extent, complex. We divide these adding commits into four kinds on the basis of this finding: *simple adding*, *adding function*, *adding class*, and *other*. We give their definitions in detail as follows:

**Simple adding**: This refers to adding several lines of codes which have no logical relationship as shown in figure 6.

**Adding function**: This refers to adding a new function aiming at fulfilling a specific functionality as shown in figure 7.

```
1 package.json                                                              View
@@ -4,6 +4,7 @@
4   4      "version": "1.0.0",
5   5      "description": "A JavaScript library that provides a simple API, and powerful features to make working with audio a breeze. Easil
6   6      "main": "soundjs.js",
    7    + "jsdelivr": "lib/soundjs.min.js",
7   8      "directories": {
8   9        "doc": "docs",
9   10       "example": "examples",
```

**FIGURE 6.** Simple adding

```
+     union(){
+for(i=[0:1:33]){
+     hull(){
+     translate([xarray[i],yarray[i],0])//cylinder(r=.5,h=10);
+         sphere(3.5,$fn=10);
+     translate([xarray[i+1],yarray[i+1],0])//cylinder(r=.5,h=10);
+         sphere(3.5,$fn=10);
+
+         }}
```

**FIGURE 7.** Adding function. A new function named union is added

**Adding class**: This refers to adding a new class when finishing a target functionality, including giving definitions to the variables and functions in this class. A case is shown in figure 8.

```
+public class UserController {
+
+    @Autowired
+    UserRepositoryImpl userRepository;
+
+    @PostMapping
+    @ResponseStatus(HttpStatus.CREATED)
+    public User register(@RequestBody UserEntity user) {
+        return userRepository.save(user);
+    }
+
+    @GetMapping
+    public User get(@RequestParam("id") int id) {
+        return userRepository.findById(id);
+    }
+}
```

**FIGURE 8.** Adding class. A class named UserController is added

**Other**: Anything that does not belong to the above conditions belongs to this category and it mainly represents two conditions. One is adding several statements with inner logical relationship such as an if conditional branch shown in figure 9. Due to the logical relationship, we do not consider this condition simple. Another condition is modifying values or types of some variables in the program as shown in figure 10.

```
+if (query.indexOf('http') !== -1) {
+    opts.url = argv._[0];
+}
+
```

**FIGURE 9.** Adding an if conditional branch

```
-       final Iterator<E> iterator = Collections.singleton(element).iterator();
+       final Iterator<E> iterator = elements.iterator();
```

**FIGURE 10.** Modifying the value of a variable

According to our definition, organ refers to code related to some specific functionalities, meaning that an organ must be practical adding. As a result, unpractical adding like the situations in Figure 2 to Figure 5 are not in the scope of our consideration when talking about organs. We calculate the number of occurrences of each kind in each project, and results are given in Table IV.

Adding function is the most widely used kind of commits with its percentage reaching 35.91% while adding class is the most uncommon kind. That is probably because adding a new function is direct and easy to conduct while adding a new class into the project may break the structure of this project. Hence, engineers prefer to use a new function and try to avoid creating a new class.

**Answer to RQ2:**

Organs from *adding commits* present four kinds of contents: simple statements, a new function, a class, and some other conditions, proving the diversity of the organs. Therefore, engineers can select from multiple options when they need to extract some functionalities from the repository.

TABLE IV
NUMBER OF OCCURRENCES OF EACH KIND OF ORGAN IN THE 12 PROJECTS AND THEIR PERCENTAGES

| **Project Name** | *simple adding* | *adding function* | *adding class* | *other* | **Total amount of *practical adding*** |
|---|---|---|---|---|---|
| SoundJS | 9 | 31 | 11 | 11 | 62 |
| dicisiontree | 4 | 0 | 3 | 2 | 9 |
| awesome-nlp | 21 | 3 | 4 | 5 | 33 |
| Stethoscope | 3 | 7 | 7 | 2 | 19 |
| polygons | 1 | 5 | 0 | 1 | 7 |
| lightcrawler | 0 | 1 | 0 | 1 | 2 |
| a11y | 10 | 8 | 6 | 11 | 35 |
| lint | 20 | 9 | 0 | 11 | 40 |
| square.github.io | 3 | 4 | 1 | 8 | 16 |
| twitter.github.com | 3 | 5 | 1 | 2 | 11 |
| requery | 16 | 48 | 24 | 14 | 102 |
| FabulousFilter | 0 | 0 | 1 | 0 | 1 |
| **In total** | 90 | 121 | 58 | 68 | 337 |
| **Percentage(%)** | 26.71 | 35.91 | 17.21 | 20.17 | 100.00 |

### C. RQ3: WHAT IS THE PORTION OF EASY-TO-TRANSPLANT ORGANS?

In this section, we first define the criteria to judge whether an organ is easy-to-transplant or difficult-to-transplant. Previous works on automated program repair [33, 34, 35] prefer to use lines of code (LoC) to measure complexity degree. However, this criterion is not suitable for organ transplantation in our mind: an organ is convenient for transplantation if it is independent even if it has thousands of lines of code. Thus, the criteria should be based on the degree of relevance of the added codes with the source codes. An easy-to-transplant organ should possess no variables defined in the context, which means that it presents weak relevance with the source codes. Thus, this organ can be transplanted into another program directly and conveniently. On the contrary, difficult-to-transplant organs display strong relevance with the source codes. Consequently, they may use the variables, functions, or classes defined in the source codes and need special operation when being transplanted. *Unpractical adding* is not in the scope of our consideration. We calculate the number of occurrences of each kind of organ and their percentages in Table V.

**Answer to RQ3:**

According to Table V, approximately 70% of the practical adding commits are easy-to-transplant organs, indicating that most organs in the repository can be extracted conveniently and thus our topic is meaningful.

TABLE V
NUMBER OF OCCURRENCES OF EASY-TO-TRANSPLANT AND DIFFICULT-TO-TRANSPLANT ORGANS IN EACH PROJECT AND THEIR PERCENTAGES

| **Project Name** | *easy-to-transplant* | *difficult-to-transplant* | **Total amount of *practical adding*** |
|---|---|---|---|
| SoundJS | 34 | 28 | 62 |
| dicisiontree | 5 | 4 | 9 |
| awesome-nlp | 30 | 3 | 33 |
| Stethoscope | 4 | 15 | 19 |
| polygons | 5 | 2 | 7 |
| lightcrawler | 1 | 1 | 2 |
| a11y | 26 | 9 | 35 |
| lint | 30 | 10 | 40 |
| square.github.io | 14 | 2 | 16 |
| twitter.github.com | 6 | 5 | 11 |
| requery | 68 | 34 | 102 |
| FabulousFilter | 1 | 0 | 1 |
| **In total** | 224 | 113 | 337 |
| **Percentage(%)** | 66.47 | 33.53 | 100.00 |

### D. RQ4: HOW TO TRANSPLANT THESE ORGANS?

In this section, we propose our manual transplantation strategy after identifying where the organs exist and introduce it in detail.

Methods for different organs should differ according to whether the organs are easy-to-transplant or difficult-to-transplant. Easy-to-transplant organs can be transplanted directly, difficult-to-transplant organs, however, need special operation for the undefined variables. Based on this thinking, the method should contain a judgement step and take operation based on different results. We first extract the codes added by developers and then conduct the check. We then take different policies for different types of organs. The framework of our method is shown in Figure 11. What need special attention is that the "variable" we discuss in the next is in the broad sense. It may refer to a real variable in the program, a function that is called, or a class, anything that appears in the added codes.

**Added code extraction:** In the first step of our method, we aim to extract the codes added by the contributors in the repository by finding out corresponding codes from adding commits. These codes are in green font and there is a "+" before each line, indicating that this step is easy to conduct. What needs special attention is the integrity of the extraction when the added codes appear in several places and does not appear as a whole.

**Code check:** In this step, we check the types of these organs and make classification. Follow-up processing is based on the classification result, making this step crucial. We list all the variables in the organ and check whether they are defined in these codes. There are two cases of results: one is that all the variables are defined in this organ, which suggests that this organ has weak relevance with the external part and thus can be transplanted without any other operation. We go to the transplantation step directly under this condition. Another condition is that there exists variables being undefined in this organ, which means these variables are defined in the original codes but are called here, suggesting that this organ has strong relevance with the external part. Hence, we need to deal with these undefined statements, otherwise they may introduce some unpredictable errors after being transplanted directly. We go to the related code extraction step to take special operation under this condition.

**Related code extraction:** In this step, we aim to record all the related statements about the variables which are defined in the original codes but called in the organ. There are two kinds of variables being under our consideration. One is local variables in the function in which the organ is added. We record all the statements which modify the values of these variables before where the organ is added on this occasion. Another is global variables in the class. We enlarge our search space into the whole class under this condition. We build call graph of the functions in this class using the method introduced in [41] and record corresponding statements before where the organ is added in the graph. We record the related statements according to their order of appearance in the call graph and this process is executed until all the statements

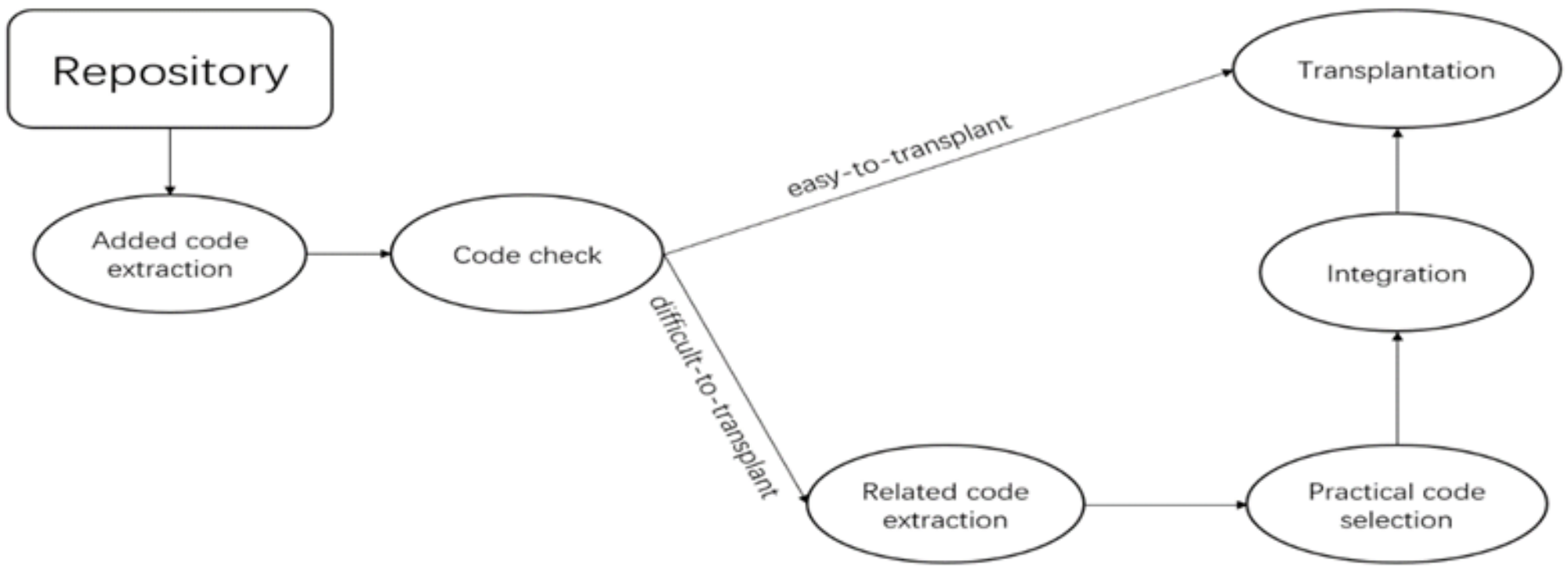

FIGURE 11. **The framework of our methodology**

about each undefined variable have been recorded.

**Practical code selection:** We have recorded all the statements related to the variables in the previous step. In this step, our goal is to select out practical statements from these statements we have recorded. Either local variables or global variables may be changed for several times before being called in the organ. Thus, what we only need to do is to reserve the latest changes of these variables, record the newest values of these variables before their being called, and delete the statements before. An example of this is shown in Figure 12. Suppose line 10 to 13 is the organ added by the developers and the variable a is defined in the source codes. During the previous step, we select out all the statements about a, they are line 1, line 3 to 6, and line 8. The initial value of a is 10. It turns to 0 after the loop from line 3 to 6 and then it is changed to 10 again. We preserve line 8 in this step in that it defines the latest value of the variable a. What needs special attention is that while we record the latest values of the variables, we also need to record their types. As a result, the final practical codes selected out from our example is int a = 10.

After this processing, we record the latest values and types of the undefined variables in the organ. These statements are called a vein to the organ according to [13] which means they have some relationship in functionality and must be together when being transplanted.

```
1.  int a = 10;
2.  int b = 0;
3.  for(; a>0; a=a-1)
4.  {
5.      b++;
6.  }
7.
8.  a = 10;
9.
10. if(a > 5)
11. {
12.         System.out.print("Just an example");
13. }
```

FIGURE 12. **An example for practical code selection**

**Integration:** In this step, we combine the vein and the organ by adding the statements we have recorded into the front of the organ. Since the vein gives definition to all the undefined variables in the organ, the entirety has no grammatical errors and can be used for transplantation. The entirety in our example is shown in Figure 13.

```
1. int a = 10;
2. if(a > 5)
3. {
4.         System.out.print("Just an example");
5. }
```

FIGURE 13. **The entirety in our example**

**Transplantation:** In our last step, we transplant the organ into the host, the project which will receive the organ. A crucial problem here is how to get the suitable location for the organ. We aim to transplant the organ into the class which has similar functionality with where the organ comes from and a fitness function is used to solve this problem. We record the name of the class where the organ is extracted from as O before transplanting and then write down all the host's class names in a set called C. It is observed that the name of the class represents its function in most cases. For example, the class with the name of *Backupdata* often intends to backup data. It is where our intuition comes from: we can assess the similarities in function by measuring the similarities in names. We split the class name into several single words and the fitness function is defined as follows:

$$f(x,y)= (N(x,y))/(N(x)+N(y)) \quad (1)$$

where x, y represent names of the two classes under measurement, N(x,y) represents the number of words that are commonly contained in the names of the two classes, and N(x) and N(y) represent the numbers of single word in the name of x and y, respectively. Every time we input two names into this fitness function, it can return the similarity value between these two names.

The fitness function is used to measure the similarity between O and each name in C and it returns a list according to the similarities. We choose the class whose name is most similar to O each time and its function is also most similar to O's according to our hypothesis, indicating that it is the suitable location for the organ. We then transplant the organ into this class and check whether the host shows specific features that we want it to possess. This process ends when the host passes the test (a successful transplantation) or all the classes have been tried (a failed transplantation).

This section proposes a method for manually extracting organs from GitHub repository and transplanting them into the host. In the next section, we do some manual experiments to check the practicality of our strategy.

### E. RQ5: DO THE ORGANS EXTRACTED FROM GITHUB REPOSITORY WORK WELL?

In this section, we aim to check the performances of the organs extracted from GitHub repository. We first manually extracted thirty organs in three different programming languages (Java, Python, and C) utilizing our strategy and then made unit tests for them using four different tools. We then transplanted three organs written in Java into a specific platform to check whether the system possessed the function we would like it to have after transplantation. We describe the process and results of the experiment in detail in the following.

#### 1) UNIT TESTS

In this section, we extracted organs written in three different programming languages and made unit tests for them employing four different tools, aiming at checking their performances in reliability and integrity in functionality.

Organs were chosen on the criteria that it must produce a single output value for our convenience to make judgement. Hence, the organs we extracted are all about mathematic functionalities.

**Java:**

The Apache Commons Math project [5] is a library of lightweight, self-contained mathematics components addressing the common practical problems in the Java programming language. We extracted 10 organs from its repository, each of which completes a mathematic functionality and follows our criteria.

JUnit test [27] is a widely used method around the world in writing tests for solving real-world problems in programs written in Java. In our experiment, we applied this test framework for our testing. As Java is one of the most popular programming language in the world and many unit test techniques have been well studied for it, we chose to utilize two tools to conduct unit tests for Java organs (one can generate test cases automatically, another need manual defined test oracles).

EvoSuite [28] is a commonly used tool for automated test suite generation, combing SBST (search-based software testing) [29] with DSE (dynamic symbolic execution) [30]. Empirically, it can increase code coverage up to 63% [28]. This tool brings great convenience for our testing since the input values and test oracles are generated automatically. We utilized this tool for generating JUnit test cases for organs we extracted, conducted unit test, and calculated passing rate for each organ. The results are illustrated in Table VI.

It is shown in the results that all the organs have high passing rates up to 60% and seven of them (except from No.1, No.2, and No.4) reach 100%. The average passing rate exceeds 90% for all the organs with more than 2,600 lines of code and over 190 test cases.

The number of test cases generated for organ 1 is not the same with that of executable cases, meaning that some of the test cases cannot be executed, as some protected methods called in the test cases cannot be visited outside of the class it belongs to. Thus, 5 test cases generated by EvoSuite lost their effectiveness. Therefore, we ignore these 5 test cases when calculating passing rate.

JUnit Test Generator [6] is another tool for unit test which can generate the whole test framework for all the functions in the class automatically. Since all needed instances are initialized as null when using this tool, we manually designed the inputs and test oracles for each test case based on efficacy of each function. Test cases generated by this tool obey the naming rule that the name of each test case is corresponding to the function tested by it, making this procedure simple to conduct. The results are shown in Table VII.

Generally speaking, this tool generates 125 test cases, a reduction of nearly a half in amount, compared with those generated with EvoSuite. This may be caused by the different test granularity of these two tools: EvoSuite works at a fine-grained level while JUnit Test Generator works at a coarse-grained level. For example, in the testing process of Organ 2 (Evaluation RmsChecker), JUnit Test Generator just produced one test case to check the function named Converged, while EvoSuite not only produced two test cases generating one right input and one wrong input and using *assertTrue()* and *assertFalse()* to check this function, but also examined the *NullPointerException*. This coarse-grained performance of JUnit Test Generator is accordance with the test results: it leads to a higher passing rate than EvoSuite. The test cases are all executable and the organs seem to have a better performance here: only two of them (No.1 and No.5) failing to pass all the tests, each of them with a passing rate exceeding 90%, and the average passing rate reaching 97.6%.

Both of the tools verify the good performances in reliability of these Java organs we extracted, proving the practicality of our strategy.

**Python:**

SymPy [7] is a Python library for symbolic mathematics, possessing 655 contributors and more than 30,000 commits. We extracted 10 organs from its repository in this part. Unit test is easy to conduct in that there is a comprehensive instruction for each class in this project with input samples and output samples in all possible conditions in it. We design our input values and test oracles based on the instructions.

TABLE VI
TEST CONDITIONS OF THE 10 JAVA ORGANS BY EVOSUITE

| **Organ Number** | *function description* | *LoC (lines of code)* | *test cases* | *executable cases* | *passing cases* | **Passing rate (%)** |
|---|---|---|---|---|---|---|
| 1 | Psquare Algorithm | 980 | 31 | 26 | 16 | 61.54 |
| 2 | Evaluation RmsChecker | 58 | 4 | 4 | 3 | 75.00 |
| 3 | Runge-kutta Integrator | 43 | 1 | 1 | 1 | **100.00** |
| 4 | Ball Generation | 139 | 10 | 10 | 6 | 60.00 |
| 5 | HelloWorldExample | 169 | 8 | 8 | 8 | **100.00** |
| 6 | InsufficientDataException | 33 | 2 | 2 | 2 | **100.00** |
| 7 | Sparse Gradient | 908 | 109 | 109 | 109 | **100.00** |
| 8 | Binomial Confidence Interval | 185 | 19 | 19 | 19 | **100.00** |
| 9 | Earth Movers Distance | 28 | 4 | 4 | 4 | **100.00** |
| 10 | Midpoint Integration | 148 | 11 | 11 | 11 | **100.00** |
| **In total** | | 2,691 | 199 | 194 | 179 | 92.27 |

TABLE VII
TEST CONDITIONS OF THE 10 JAVA ORGANS BY JUNIT TEST GENERATOR

| **Organ Number** | *function description* | *LoC (lines of code)* | *test cases* | *passing cases* | **Passing rate (%)** |
|---|---|---|---|---|---|
| 1 | Psquare Algorithm | 980 | 34 | 32 | 94.12 |
| 2 | Evaluation RmsChecker | 58 | 1 | 1 | **100.00** |
| 3 | Runge-kutta Integrator | 43 | 1 | 1 | **100.00** |
| 4 | Ball Generation | 139 | 2 | 2 | **100.00** |
| 5 | HelloWorldExample | 169 | 10 | 9 | 90.00 |
| 6 | InsufficientDataException | 33 | 2 | 2 | **100.00** |
| 7 | Sparse Gradient | 908 | 67 | 67 | **100.00** |
| 8 | Binomial Confidence Interval | 185 | 5 | 5 | **100.00** |
| 9 | Earth Movers Distance | 28 | 1 | 1 | **100.00** |
| 10 | Midpoint Integration | 148 | 2 | 2 | **100.00** |
| **In total** | | 2,691 | 125 | 122 | 97.60 |

PyUnit [8] is a Python language version of JUnit. This unit testing framework uses a proven testing architecture and is easy to be used, being part of the Python 2.1 standard library. We took our unit test by utilizing this framework and the results are listed in Table VIII. It is observed from the results that organs in Python language are not as long in length as organs in Java language: these ten organs have just 1,029 lines of code in total, less than half of the lines number of Java organs. Correspondingly, only 61 test cases are created, of which 50 pass the tests, making the average passing rate surpass 80%. Organ 2 has the lowest passing rate: it passes only three of the five test cases. Four organs (No.1, No.6, No.9, and No.10) pass all the test cases, proving their great performances.

Organs in Python language shows passing rate of over 80%, though a rate not as high as that of Java organs, thereby proving their good performances in reliability.

**C:**

CML [9] is a pure-C math library with a great variety of mathematical functions. We extracted 10 organs from this repository, each of which fulfills a simple mathematical function.

LDRA Testbed [31] is a software analysis tool providing core static and dynamic engines for code written in C/C++ language. Since the initial values and test oracles need to be input manually, we designed our test cases based on the ability of the organs. This tool also provides a coverage analysis through calculating statement coverage and branch coverage by the test cases automatically. We continuously added new test cases into the test suite until both of the coverages reached 100% and recorded the corresponding test results shown in Table IX.

Although four organs (No.3, No.4, No.7, and No.9) pass all the tests, Organ 2 only has 50% passing rate and the average passing rate is 80.56%, the lowest among these three types of organs. A possible reason for this is that there are pointers existing in these organs and the pointers may cause confusion in the program structure when not being processed properly. This phenomenon indicates that how to get accurate definitions and values for pointers undefined in the organ still has a large research space.

In this section, we used four tools to make unit tests for organs written in three different languages which are Java,

TABLE VIII
TEST CONDITIONS OF THE 10 PYTHON ORGANS BY PYUNIT

| **Organ Number** | *function description* | *LoC (lines of code)* | *test cases* | *passing cases* | **Passing rate (%)** |
|---|---|---|---|---|---|
| 1 | fibonacci | 49 | 3 | 3 | **100.00** |
| 2 | bernoulli | 92 | 5 | 3 | 60.00 |
| 3 | bell | 99 | 6 | 5 | 83.33 |
| 4 | harmonic | 155 | 13 | 9 | 69.23 |
| 5 | euler | 97 | 10 | 8 | 80.00 |
| 6 | genocchi | 86 | 4 | 4 | **100.00** |
| 7 | factorial | 177 | 5 | 4 | 80.00 |
| 8 | factorial2 | 99 | 4 | 3 | 75.00 |
| 9 | subfactorial | 61 | 5 | 5 | **100.00** |
| 10 | risingfactorial | 114 | 6 | 6 | **100.00** |
| **In total** | | 1,029 | 61 | 50 | 81.97 |

TABLE IX
TEST CONDITIONS OF THE 10 C ORGANS BY TESTBED

| **Organ Number** | *function description* | *LoC (lines of code)* | *test cases* | *passing cases* | **Passing rate (%)** |
|---|---|---|---|---|---|
| 1 | median | 22 | 3 | 2 | 66.67 |
| 2 | mean | 16 | 4 | 2 | 50.00 |
| 3 | max | 23 | 3 | 3 | **100.00** |
| 4 | min | 23 | 3 | 3 | **100.00** |
| 5 | max_index | 30 | 5 | 4 | 80.00 |
| 6 | min_index | 30 | 5 | 4 | 80.00 |
| 7 | absolute deviation | 20 | 2 | 2 | **100.00** |
| 8 | variance | 128 | 6 | 5 | 83.33 |
| 9 | covariance | 59 | 2 | 2 | **100.00** |
| 10 | kurtosis | 33 | 3 | 2 | 66.67 |
| **In total** | | 384 | 36 | 29 | 80.56 |

Python, and C, respectively. Results are satisfactory with the highest passing rate reaching 97.6% and the lowest still exceeding 80%, proving that organs extracted by our strategy possess good performances in unit tests.

### 2) SPECIFIC PLATFORM

MiCode Notes is an open-source edition of XM notepad and is widely used by Android users. More information about this application is on the MiUi[13] website. This tool was chosen as our specific platform in that it is open source and free to use and possesses clear logical structure which is suitable for being a host in transplanting. In this section, we extracted three organs from GitHub repository and transplanted them into MiCode Notes to assess if they could work well. As it is the first step towards transplanting organs from open-source repository, these three organs, from three different contents which are *simple adding*, *adding class*, and *adding function*, respectively, were chosen on the criteria that no more than 20 lines of code to reduce the complexity in our manual experiment.

**Simple adding**[14] :

One commit in [1] indicates that we can add music player to the program by adding the following codes:

```
playerView = findViewById(R.id.player_view);
super.onStart();
```

We added this organ into the function *Oncreate()* in the class named *NodeEditActivity* after four attempts in that the name of the donor file (*PlayerActivity*) possessed a common keyword "activity" with the host file name and there were three host file with the same word listing before *NodeEditActivity* alphabetically. Our aim was playing music automatically when editing notes and this organ worked when we opened a new note.

**Adding class**:

We added welcome screens by defining a new class just like the following according to a commit in our last project in the dataset for RQ1 to RQ3 ([3]) :

[13] http://www.miui.com/index.html

[14] The present experimental result is recorded as a video and publicly available on https://youtu.be/bgNk2B51RD4

```
public class Fragment {
    public View onCreateView (LayoutInflater inflater, ViewGroup
  container, Bundle savedInstanceState) {
      super.onCreateView(inflater, container, savedInstanceState);
    View view = inflater.inflate(R.layout.fragment, container, false);
    return view;
    }
};
```

We transplanted this organ into *WelcomeActivity* class and modified some of the variables into ours. This time, the name of original donor file was *FragmentExampleActivity* and we tried for five times. Subsequently, we opened the software again and determined that the new functionality worked well. The screenshot is displayed in Figure 6a.

**Adding function**:

We added a gesture password into this application to protect the privacy of users. One commit in [2] provides a method by adding the following function:

```
public void startLockActivity() {
  Toast.makeText(this, "set password", Toast.LENGTH_SHORT).show();
  Intent intent=new Intent();
  startActivity(intent);
}.
```

We added this function to *NoteListActivity* class. This period was the longest one among three experiments since the donor file name *EncryptAndDecryptController* had no common keyword with our target host file and thus we had to try these files in host system one by one. In total, fifty attempts were conducted. We also added necessary materials like background pictures and ran this program again. Figure 6b illustrates the screenshot of setting gesture password. This result indicates that we successfully add a password before logging in.

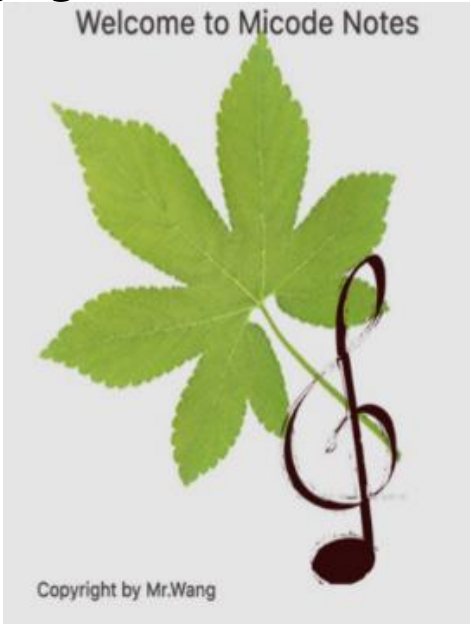

**FIGURE 14a.** Welcome screen

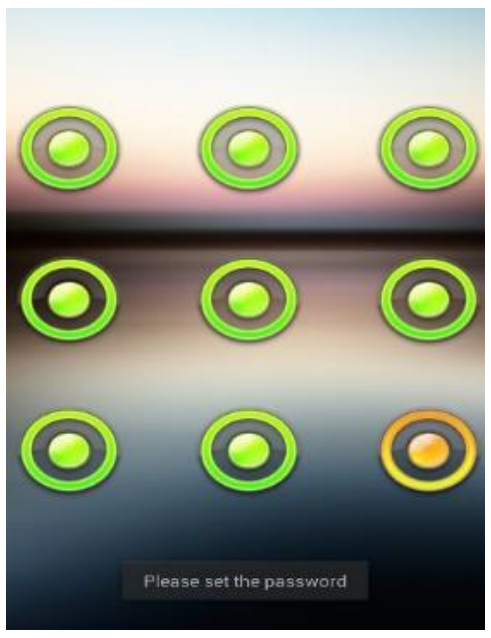

**FIGURE 14b.** Screen of setting password

We count the number of three types of atomic operations after transplanting organs into the host in each experiment to assess the extra workload after transplantation and to judge the performance of our transplantation way: adding a statement, deleting a statement, and modifying a statement. Adding is used for some materials which are needed, to add new items in the xml file, or to import the dependency we need. For example, when we transplanted the first organ, we needed to import the class named *com.google.android.exoplayer2* to make the host runnable while we also needed to add two new items in note-list.xml for the third organ. Deleting occurs when some statements cause errors due to the repetition problem. Modifying refers to making little change in a certain statement such as changing the variables' names into the hosts' or changing the configuration file. An instance is in the process of transplanting the second organ, we needed to change the enter point of the whole program to the *WelcomeActivity* class and this was conducted in *AndroidManifest.xml*. Results are shown in Table X, demonstrating that the most complex one contains only 13 atomic operations which are 10 adding operations, 1 deleting operation, and 2 modifying statements. It is shown that the extra work after transplantation is little and the only operations have no relation with the program structure, proving the ability of our transplantation strategy.

TABLE X
NUMBER OF EACH TYPE OF ATOMIC OPERATION IN EACH TRANSPLANTATION EXPERIMENT

| Form of organ | adding | deleting | modifying |
|---|---|---|---|
| Simple adding | 1 | 0 | 1 |
| Adding class | 3 | 0 | 1 |
| Adding function | 10 | 1 | 2 |

**Answer to RQ5:**

We first make unit tests for thirty organs extracted by our strategy from GitHub repository covering Java, Python, and C programming languages. All the organs have good performances in the tests with at least 80% of the test cases are passed for each language. We then extract three Java organs in different forms and transplant them into our experiment platform. All of them work well after being transplanted, thereby providing three new and practical functionalities for MiCode Notes.

## IV. THREATS TO VALIDITY

Six factors may affect the accuracy of our study. The first is from our dataset for RQ1. Due to the limitation of manual analyses, the scale of our dataset is restricted and each project in our dataset may not be representative since GitHub repository provides several criteria like *most folks* and possessing more collectors may not mean possessing more contents in the repository. Meanwhile, GitHub repository is constantly changing: contributors from all over the world are constantly developing new features on this platform. Thus, many commits become available after our data collection. Consequently, our dataset suffers from a limited scale and may not represent the latest situation. Second is that we may remove several organs by only concerning the adding commits. According to the definition, organs refer to some codes with specific functions. Thus, commits in the repository with keywords, such as "fix" or "update", may also provide practical organs. Considering adding commits only may influence the coverage on potential organs in the repository. Third is the inappropriate style of comments. Many comments do not include the keywords we mentioned. Hence, potential organs may not be assigned to adding commits and may be missed out. Many commits in polygons are assigned to other category due to this reason. Fourth is the coverage of programming language in our experiment. We make unit tests for organs in Java, Python, and C languages and only transplant Java organs. Although these languages are three of the most presentative languages nowadays, there are still a large number of other programming languages in the open-source communities. Whether our strategy is suitable for other languages still need to be studied. Fifth is the capabilities of our test cases to find out defects. We manually design most of our test cases since only one of the four testing tools we used can generate test suite automatically. The quality of these manually created test cases may not be that high and as a result, the performances of organs we extracted may not be that good. The last one is we just make unit tests for organs with mathematic functions for the convenience of the tests. However, there are many other functions the organs possess in the real world. Whether these organs still have good performances in unit test need to be checked.

## V. RELATED WORK

Code transplantation is once used for automated repair. GenProg (*Genetic Programming*) and RSRepair (*Random Search Repair*) [10,12] transplant code from one location to another to fix bugs, promoting the development of automated program repair. CodePhage [36] is designed for transferring codes from one suitable application to another to repair errors. Harmen et al. first introduced the concept of organ in [13]. He used five-donor programs and three-host programs to conduct 15 transplantation experiments. He then conducted an extensive experiment in Kate by utilizing this tool in [14]. Based on this work, CCC (*CodeCarbonCopy*) [15] supports data representation translation and works well on eight transfers between six applications. All the experimental results are satisfying, but the transplanted organs are specific and have been prepared well. The problem of organ richness has not been solved and our study aims at solving this severe problem by mining open-source information.

Open-source communities contain a large number of resources, which can provide services for software. Yuan et al. [17] utilize an open-source software in characterizing logging practices. Moreover, Coelho et al. [18] presented an empirical study about bug hazards extracted from more than 600 open-source Android projects and called for tool support to help developers understand exception handling. Xiong et al. [19] introduces the open-source approach into the field of automatic program repair. They analyze the content of the question and answer section of the open source community, extract the pre-repair and repair methods, and then match the content of the program to be repaired based on the abstract syntax tree to get the appropriate repair method. *ssFix* [32] performs syntactic code search to find existing code from an open-source database and further leverages such code to produce patches for bug repair.

There has also been a lot of research concerning on reuse in open-source communities. Li et al. [16] proposed a method to recommend existing codes to developers for reuse based on software term database. Another tool named *Code Conjurer* [20] aimed at making codes available for users with almost no effort. Components are the main building blocks for software architectures and component reuse has been studied for a long time such as how to address the problem of component identification from object-oriented software in [21,24], how to mine components from similar software in [22], and an improved method of identifying components in [23]. However, components do not have the integrity on functionality and are not as convenient as organs when being reused since they only emphasize on simple metrics like high cohesion and loose coupling and thus may not lead to the identification of practical functionalities. Previous studies have found and discussed this weak point in [23,25]. Code clone, which refers to reusing some code fragments by copying with or without minor modifications, has been classified into four types based on both the textual and functional similarities [37]. A previous study [38] shows that code clone is harmful in software maintenance and evolution, thus many approaches have been proposed to detect the clones, such as [39] and [40]. To some extent, code clone is similar to organ transplantation but organ transplantation is a finer-grained conception since organ emphasizes on the feature of interest while the cloned code may contain redundant statements.

To the best of our knowledge, this study is the first to propose a strategy for extracting organs from open-source repository, utilizing open-source communities to make the first exploration relating to organ extraction and transplantation based on the large-scale dataset.

## VI. CONCLUSION AND FUTURE WORK

In this paper, we presented an empirical study on extracting organs from GitHub repository for transplantation. We first analyzed statistics from 12 GitHub projects. We found that practical organs are abundant in commits with “add” as a key word in comments, organs in *adding commits* totally possess four kinds of contents, and nearly 70% of the organs are easy-to-transplant. We then put forwarded our strategy for transplantation and manually extracted 30 organs in three different programming languages (Java, Python, and C) to make unit tests for them using four different testing tools. Results show that all these organs have good performances in unit tests with the lowest passing rate among them still passing 80% and the highest one reaching 97%. We transplanted three Java organs into a specific platform to check whether they can work well in the new system in the last. Results demonstrate that these organs work well after being transplanted, providing three new functionalities for our experiment platform. It can be concluded that our transplantation strategy is feasible and it really exists a large number of practical organs in the open-source repository.

During our manual experiment, we find that there are two main problems that need to be solved in automatically extracting organs from open-source repositories. First is that contributors may not only add some codes but also delete some original codes. This condition is shown in Figure 15. It may indicate that the similar deletion need to be conducted in the host system if the host and donor have same structures. Here we cannot just transplant the codes being added into the new system, otherwise, duplication will occur in the host system. This process will be a great challenge for automated transplantation. Second is that codes are added into different files in many organs, which means several files complete this function together. Figure 16 is an example of this condition. On this occasion, it is hard to determine where to transplant these codes in. The use of names to find similarity between donor and host file can lead to organs being wrongly transplanted and sometimes it is in low-efficiency (like *adding function*). Thus, how to find the suitable places for organs when transplanting is a very valuable research direction. Also, we aim to extract and transplant more potential organs from commits with different keywords like "fix" and "modify". The distribution situation of organs in GitHub repository needs detailed investigation.

This study may provide new ideas about software reuse. In the future, we will first choose some other tools for generating test suite and make tests for the organs written in other languages to check the universality of our strategy. Then we will transplant more organs and subsequently develop a tool for automated extraction of organs from repository by summarizing the experience from manual experiment. Organ transplantation will not only benefit software reuse but also provide convenience for program repair. Thus, we believe that our work is highly significant.

```
-var rootSubject = pkix.Name{
-        Organization: []string{"mkcert development CA"},
+var userAndHostname string
+
```

**FIGURE 15.** A case of adding new codes while deleting

```
+func (m *mkcert) uninstall() {
+        m.uninstallPlatform()
+        log.Print("The local CA is now uninstalled from the system trust store! 😮\n\n")
+}
+
 func (m *mkcert) check() bool {
        if m.ignoreCheckFailure {
                return true

ststore_darwin.go

@@ -95,6 +95,12 @@ func (m *mkcert) installPlatform() {
        fatalIfCmdErr(err, "security trust-settings-import", out)
 }

+func (m *mkcert) uninstallPlatform() {
+        cmd := exec.Command("sudo", "security", "remove-trusted-cert", "-d", filepath.Join
+        out, err := cmd.CombinedOutput()
+        fatalIfCmdErr(err, "security remove-trusted-cert", out)
+}
+
```

**FIGURE 16.** A case of codes added into different files

## ACKNOWLEDGEMENT

Now the experiment platform in the second part of RQ5 is publicly-available to support further research on transplantation and provide a convenient environment for other researchers: https://github.com/Kaka727/MiCode-Note.